\documentclass[letterpaper, 10 pt, conference]{ieeeconf}  % Comment this line out if you need a4paper

\IEEEoverridecommandlockouts                              %

\usepackage{romannum}
\usepackage{cite}
\usepackage{amsmath}
\usepackage{graphicx}
\usepackage{caption}
\usepackage{subcaption}
\usepackage{float}
\usepackage{algorithm}
\usepackage{algorithmic}
\usepackage{xcolor}
\usepackage{amssymb}
\usepackage{textcomp}
\usepackage[utf8]{inputenc}
\usepackage{hyperref}
\usepackage{comment}

\DeclareMathOperator{\E}{\mathbb{E}}

\newtheorem{remark}{Remark}

\hypersetup{
    colorlinks=true,
    linkcolor=blue,
    filecolor=magenta,      
    urlcolor=cyan,
}

\title{\LARGE \bf
Optimal Lighting Control in Greenhouses Using Bayesian Neural Networks for Sunlight Prediction*}

\author{Shirin Afzali$^{1}$, Yajie Bao$^{1}$, Marc W. van Iersel$^{2}$, and Javad Mohammadpour Velni$^{1}$% <-this % stops a space
\thanks{*This work was financially supported by USDA-NIFA-SCRI award number \#2018-51181-28365, project Lighting Approaches to Maximize Profits.}% <-this % stops a space

\thanks{$^{1}$Shirin Afzali, Yajie Bao, and Javad Mohammadpour Velni are with the School of Electrical \& Computer Engineering, University of Georgia, Athens, GA 30602, USA; {\tt\small shirin.afzali@uga.edu}, {\tt\small yajie.bao@uga.edu}, {\tt\small javadm@uga.edu}.}%
        
\thanks{$^{2}$Marc W. van Iersel is with the Department of Horticulture, University of Georgia, Athens, GA 30602, USA; {\tt\small mvanier@uga.edu}.}%} 
}

\begin{document}

\maketitle
\thispagestyle{empty}
\pagestyle{empty}

%%%%%%%%%%%%%%%%%%%%%%%%%%%%%%%%%%%%%%%%%%%%%%%%%%%%%%%%%%%%%%%%%%%%%%%%%%%%%%%%
\begin{abstract}

Controlling the environmental parameters, including light in greenhouses, increases the crop yield; however, the electricity cost of supplemental lighting can be high. Therefore, the importance of applying cost-effective lighting methods arises. In this paper, an optimal supplemental lighting control approach is developed considering a variational inference Bayesian Neural Network (BNN) model for sunlight prediction. The predictive model is validated through testing the model on the historical solar data of a site located at North Carolina ($R^{2}$=0.9971,\,RMSE=1.8\%). The proposed lighting approach is shown to minimize electricity cost by considering the BNN-based sunlight prediction, plant light needs, and variable electricity pricing when solving the underlying optimization problem. For evaluation, the new strategy is compared to: 1) a Markov-based prediction method, which solves the same optimization problem, assuming a Markov model for sunlight prediction; 2) a heuristic method which aims to supply a fixed amount of light. Simulation studies are conducted to examine the electricity cost improvements of the BNN-based approach. The results show that the BNN-based approach reduces cost by (on average) 2.27\% and 43.91\% compared to the Markov prediction-based method and the heuristic method, respectively, throughout a year.  

\end{abstract}

%%%%%%%%%%%%%%%%%%%%%%%%%%%%%%%%%%%%%%%%%%%%%%%%%%%%%%%%%%%%%%%%%%%%%%%%%%%%%%%%
\section{Introduction}

\noindent The use of supplemental lighting is an effective way for improving plant growth in controlled environment agriculture (CEA), and in particular, greenhouses. However, the electricity needed for supplemental lighting accounts for up to 30\% of the operating costs of a greenhouse \cite{van2017adaptive}. As a result, improving the cost-effectiveness of lighting plays an important role in CEA industry. Among various lighting types, light-emitting diodes (LEDs) have proven to be more effective, due to their dimming capability which allows controller to adjust the intensity of output light to any continuous level \cite{narra2004effective}. On the other hand, high-intensity discharge (HID) lamps can be dimmed only into  some discrete levels, thereby not being chosen as the primary choice for optimal lighting strategies.

Rule-based supplemental lighting control methods have been proposed since a long time ago; in \cite{carrier1994description}, a rule-based system for HID lamps (with on/off control) was developed to achieve a pre-defined amount of daily light integral (DLI) within a specified photoperiod for tomato. DLI (with the unit of $\,mol\,m^{-2}\,d^{-1}$) is the sum of photosynthetic photon flux densities (PPFDs) over 24 hours; PPFD is the light-photon numbers in the photosynthetically-active range (400-700 nm) per square meter per second with the unit of $\mu \,mol\,m^{-2}\,s^{-1}$. 
 
Another rule-based control method is LASSI (light and shade system implementation) \cite{albright2000controlling} which uses on/off control of HID lamps in combination with movable shades and a sunlight prediction to achieve a consistent DLI. A later version of LASSI used an improved sunlight prediction, and it was demonstrated that using a better prediction could improve the accuracy of lighting control \cite{seginer2006}. Moreover, DynaLight system in \cite{maersk2011software} uses an on/off control of HID lamps, considering variable electricity pricing, weather forecast and a leaf photosynthesis model to reduce electricity cost subject to reaching a minimum required daily sum of photosynthesis. However, this system is provided with weather forecasts twice daily in hourly resolution \cite{maersk2011software}, which is not precise enough for true optimization. 

Adaptive lighting method \cite{pinho2013dynamic} is another rule-based control approach in which cost of lighting is reduced by adjusting the duty cycle of LEDs to reach a specific threshold based on PPFD levels. In previous research studies on supplemental lighting control with LEDs, it was assumed that either there is complete information about daily sunlight intensity \cite{weaver2019photochemistry}, or real-time information of sunlight intensity is used to provide fixed levels of PPFD to greenhouse plants \cite{schwend2016}. However, both of these approaches have shortcomings because future sunlight intensity is unknown, and controlling supplemental lighting based solely on the current PPFD does not account for DLI or daily photosynthesis. 

In our prior studies \cite{mosharafian2021optimal,afzali2021optimal}, we developed optimal supplemental lighting strategies for both HID lamps and LEDs in greenhouses, which significantly reduce electricity cost. We formulated the supplemental lighting control problem as
a constrained optimization problem, and we aimed at minimizing electricity cost of supplemental lighting, while considering sunlight prediction, plant light needs, and variable
electricity pricing in our model. This strategy predicts future sunlight intensities based on a Markov model. For LEDs, the optimization problem is continuous \cite{mosharafian2021optimal}; however, for HID lamps, we ended up with a discrete constrained optimization problem which was solved using the method of multipliers and a reinforcement learning (RL) algorithm \cite{afzali2021optimal}. 
Although our lighting strategy is optimal, the accuracy of control is dependent on the accuracy of sunlight prediction (used a Markov predictive model). In the present paper, we develop our strategy for greenhouses equipped with LEDs based on a Bayesian Neural Network (BNN) model which provides more accurate predictions than a Markov model. In particular, we predict future sunlight intensities using this BNN model and solve a constrained optimization problem. 

BNN provides an ensemble of models by assigning weights probability distributions and use model averaging to enhance the prediction accuracy. Using BNN for predicting solar irradiation has been studied in previous studies \cite{lopez2005selection,yacef2012prediction}.
Researchers in \cite{yacef2012prediction} conducted a comparative study among Bayesian Neural Network (BNN), classical Neural Network (NN) and empirical models for estimating solar irradiance, and BNN outperformed the other methods ($R^{2}$=0.9715,\,RMSE=8.41\%). They considered air temperature, relative humidity, sunshine duration and extraterrestrial irradiation as their model inputs. In our research paper, we consider time (hour) and solar irradiance as the input parameters and use variational inference to build our model with a high accuracy.

\begin{comment}
model the uncertainty  
\end{comment}

The main contributions of this work are as follows: (1) we propose a variational inference BNN model with a high accuracy to predict sunlight intensities during a day; (2) using this prediction, we devise an optimal lighting strategy to minimize the supplemental lighting electricity cost.

\begin{comment}
(2) assuming only the mean for the BNN-based model, we devise an optimal lighting  strategy by solving a deterministic optimization problem; (3) assuming standard deviation other than mean for the sunlight predictive model, we enhance the lighting strategy by solving a stochastic optimization problem.
\end{comment}
 
The remainder of this paper is organized as follows. In Section \Romannum{2}, the optimization problem for supplemental lighting is introduced. BNN and sunlight prediction are discussed in Section \Romannum{3}. The proposed lighting approach is explained in Section \Romannum{4}. Simulation results are discussed in Section \Romannum{5}, and Section \Romannum{6} concludes this paper.

%%%%%%%%%%%%%%%%%%%%%%%%%%%%%%%%%%%%%%%%%%%%%%%%%%%%%%%%%%

\section{Preliminaries and Problem Formulation}

\noindent Two important parameters in plant photosynthesis are electron transport rate (ETR) and PPFD. ETR is defined as the number of electrons transported through photosystem II per square meter leaf area per second (with the unit of $\mu\,mol\,m^{-2}\,s^{-1}$). The sum of ETRs over 24-hour period is called the daily photochemical integral (DPI) with the unit of $\,mol\,m^{-2}\, d^{-1}$. Past studies (e.g., see \cite{weaver2019photochemistry}) show that sunlight power ($Wm^{-2}$) can be converted to PPFD using a conversion factor of %$4.48\times 0.45$ 
$\approx 2.02$. Furthermore, ETR and PPFD generally have an exponential relation.  Authors in \cite{weaver2019photochemistry} derived a relationship to calculate ETR with a given PPFD as

\vspace{-15pt}

\begin{equation*} \label{ppfdetr}
    ETR=a(1-e^{-k\times PPFD}),
\end{equation*}
\vspace{-15pt}

\noindent where $a$ is the asymptote of ETR, and $k$ is the initial slope of ETR divided by $a$. For instance, for ‘Green Towers’ lettuce, $a=121\: \mu \,mol\, m^{-2}\,s^{-1}$ and $k=0.00277$ \cite{weaver2019photochemistry}.

A minimum DLI is recommended for many greenhouse crops to guarantee sufficient growth and high quality production. In some cases, a specific photoperiod (time that plants are exposed to light in 24 hours) must also be achieved. However, the DPI and plant growth depend on the combination of DLI and photoperiod; longer photoperiods with the same DLI result in a higher DPI and more biomass \cite{elkins2020longer, weaver2020longer}. Therefore, DPI is a better predictor of plant growth and better suited for lighting optimization algorithms than DLI. ‘Green Towers’ lettuce needs a DPI of 3 $\,mol\, m^{-2}\,d^{-1}$ corresponding to a DLI of 17 $\,mol\,m^{-2} \,d^{-1}$ under ambient sunlight conditions \cite{weaver2019photochemistry}.

The optimization problem is formulated to minimize the total amount of light provided from supplemental lights to reach a specified DPI within a specified photoperiod. The formulation was first presented in our prior study \cite{mosharafian2021optimal}, where Markov chain was used to predict sunlight intensities during a day. Here, we leverage BNN to predict sunlight intensities and show how a more accurate predictive model can improve the lighting strategy. The optimization problem considered here is as follows:

\begin{equation} \label{optprob}
\begin{aligned}
\underset{\overline{x}}{\text{min}}\,f(\overline{x})=
\sum_{t=1}^{T} \frac{C_t}{k}\Big[\ln(\frac{a}{a-\overline{x}_t-\overline{s}_t})-s_t\Big] \\
\text{subject to:}\;\;
\sum_{t=1}^{T}{(\overline{x}_t+\overline{s}_t)} \geq \frac{\overline{D}}{m},
\\\overline{x}_t\geq 0 \:\;;\:\;  t=1,2,\cdots,T
\\\overline{x}_t\leq \overline{U}_{LED} \:\;;\:\;  t=1,2,\cdots,T
\end{aligned}
\end{equation}
where $\overline{x}_t$ is the ETR resulting from supplemental light provided by the LEDs \color{black}at time-step $t$, $\overline{s}_t$ is the ETR resulting from sunlight, $s_t$ is the PPFD received from the Sun, $\overline{U}_{LED}$ is the maximum ETR that can be achieved with the LED light, $C_t$ is electricity price in $cent/kWh$, $\overline{D}$ is the DPI needed for the plant during its entire photoperiod, $m$ is the length of each time-step in seconds, and $T$ is the number of time steps. The first constraint in \eqref{optprob} guarantees supplying the recommended DPI to the plants, while the last two ensure the lower and upper bounds on ETR based on the minimum and maximum PPFD of the supplemental LED lights.

%%%%%%%%%%%%%%%%%%%%%%%%%%%%%%%%%%%%%%%%%%%%%%%%%%%%%%%%%%%%%

\section{Sunlight Prediction with BNNs}

\noindent Solar radiation prediction requires a model for calculating future sunlight intensities. Markov models have been useful for modeling processes that can be predicted using only current information \cite{tolver2016introduction}. In our prior study \cite{mosharafian2021optimal}, it is assumed that the future sunlight intensity is related to the current observation and independent of previous intensities, and hence, the behavior of sunlight irradiance during a day can be modeled as a Markov process. For a detailed description on sunlight prediction using Markov chains, we refer to our previous work \cite{mosharafian2021optimal}. In our present work, we use BNN to increase the accuracy of sunlight prediction in our proposed method and investigate the effects of that on electricity cost. 

BNN adds uncertainty modelling to the flexibility, scalability, and predictive performance of neural networks. Feed forward neural networks are incapable of estimating the uncertainty in predictions; however, in BNN, the weights are assigned a probability distribution instead of a single value \cite{hinton1993keeping,neal2012bayesian}. We propose to use variational inference to train BNN and learn the parameters of probability distributions \cite{blundell2015weight}. Variational inference is a machine learning method for approximating probability densities. This method posits a family of densities and then finds a member of the family that is closest to the target density \cite{blei2017variational}.

A neural network can be viewed as a probabilistic model
$P(y|x,w)$; where $x$ is the input, $y$ is the output, and $w$ is the set of weights. For regression, $y$ is a continuous variable and $P(y|x,w)$ is a Gaussian distribution \cite{blundell2015weight}. Given a set of training examples $D = (x_i,y_i)_i$, the weights can be learnt by maximum likelihood estimation (MLE):

\begin{equation} \label{MLE}
\begin{gathered}
w^{MLE}=\underset{w}{\text{argmax}}\,log\,P(D|w)={\underset{w}{\text{argmax}}\,}\sum_{i}log\,P(y_i|x_i,w).
\end{gathered}
\end{equation}

\noindent MLE can lead to severe overfitting, therefore the following maximum a posteriori (MAP) is computed which has a regularizing effect \cite{blundell2015weight}

\begin{equation} \label{MAP}
\begin{gathered}
w^{MAP}=\underset{w}{\text{argmax}}\,log\,P(w|D)\\=\underset{w}{\text{argmax}}\,[log\,P(D|w)+log\,P(w)].
\end{gathered}
\end{equation}
Variational inference approximates the true posterior by finding the parameters $\theta$ of a distribution on the weights $q(w|\theta)$ that minimizes the Kullback-Leibler (KL) divergence between $q(w|\theta)$ and the true posterior $p(w|D)$ as \cite{blundell2015weight,bao2020epistemic}
\begin{equation} \label{theta}
\begin{gathered}
\theta^{*}=
\underset{\theta}{\text{argmin}}\,KL(q(w|\theta)||P(w|D))\\ = \underset{\theta}{\text{argmin}}\,(q(w|\theta)||P(w))-\E_{q(w|\theta)}[log\,P(D|w)]\\= \underset{\theta}{\text{argmin}}\,\Big(\E_{q(w|\theta)}[log\,q(w|\theta)]\\-\E_{q(w|\theta)}[log\,P(w)]\\-\E_{q(w|\theta)}[log\,P(D|w)]\Big).
\end{gathered}
\end{equation}

\noindent The resulting cost function \eqref{theta} is known as the variational free energy \cite{blundell2015weight,bao2020epistemic}. This cost function can be approximated as follows:
\begin{equation} \label{theta_ap}
\begin{gathered}
F(D,\theta)\approx
\frac{1}{N}\sum_{i=1}^{N}\Big[log\,q(w^{(i)}|\theta)-log\,P(w^{(i)})\\-log\,P(D|w^{(i)})\Big],
\end{gathered}
\end{equation}

\noindent where ${w^{(i)}}_{i=1}^{N}$ are i.i.d. samples drawn from the variational
posterior $q(w|\theta)$ which are used to evaluate \eqref{theta_ap} during a forward-pass. A training iteration includes a forward-pass and a backward-pass. Instead of parameterizing the neural network with weights directly, we paramaterize the posterior by $w_{i,j} = \mu_{i,j} +\sigma_{i,j}\xi_{i,j}$ with parameters $\theta_{i,j}=(\mu_{i,j},\sigma_{i,j})$, where $\xi_{i,j} \sim \mathcal{N}(0,1)$ and the subscript $i,j$ denotes the $(i,j)$-th entry of a matrix (aka the reparameterization trick \cite{blundell2015weight}). During a backward-pass, backpropagation calculates the gradients of $\mu$ and $\sigma$. Therefore, we use a DenseVariational layer for the implementation which requires the aforementioned parameters. For more information on the implementation of the DenseVariational layer, please refer to \cite{tran2018bayesian}.

We collected solar data from a site in Elizabeth City, NC, from 1997 to 2012 (available at NREL database \cite{NREL})  and used the data related to 1997-2008 for training the model (twelve years) and the data from 2009-2012 (four years) for testing the model. Since radiation levels and daylight duration vary from one month to another, different models (with the same architecture but different parameters) are trained for each month. To develop the BNN model, we considered $S_t$ (the sunlight irradiance at time step $t$) and $t$ as the inputs and $S_{t+1}$ as the output of our model. The model was built in Python using Keras\footnote{ \url{https://keras.io/}} and TensorFlow\footnote{\url{https://www.tensorflow.org/}} packages. We used two Dense hidden layers, each of which contained 100 neurons with the ReLU (Rectified Linear Unit) activation function, and a DenseVariational output layer with one neuron and no activation function. For model optimization, we used Adam optimizer in Keras with a 0.0001 learning rate. Other parameters were set as default. The model was trained for 2000 epochs, and the simulation studies were conducted on a computer with a 1.8GHz CPU and 8 GB RAM.

%While training the model, the mean value ($\mu\textcolor{red}{_{\bar{s}}}$) and the standard deviation ($\sigma\textcolor{red}{_{\bar{s}}}$) of the sunlight intensity can be predicted. 

For sunlight prediction, two cases are considered: (a) before sunrise, for which the mean value of the training data is considered as the prior knowledge of sunlight irradiance for the rest of the day; and (b) during the day, in which prediction information from BNN model is used as sunlight information. Specifically, we sample $10$ models from the trained BNN and use the average of the predictions of the models as the prediction of sunlight irradiance.

%%%%%%%%%%%%%%%%%%%%%%%%%%%%%%%%%%%%%%%%%%%%%%%%%%%%%%%%%%%%%%%%%%

\section{Proposed Approach for Optimal Lighting Control}

\noindent A photoperiod of 16 hours is common for greenhouse lettuce production and used to illustrate the operation of the control strategy. The optimization problem is solved at each time step (with the length of $m$ seconds) during the allowed photoperiod for each day, and supplemental light is provided up to the optimal PPFD calculated for that time step. The process is repeated every $m$ seconds time step, for a total number of $T={16\times 3,600} \, /{m}$ when a 16-hour photoperiod is used. Rather than using actual values of future sunlight intensities, which are not available in practice, the predicted values from the BNN model are substituted in \eqref{optprob}. Consequently, \eqref{optprob} can be rewritten as
\begin{equation} \label{optprobedited}
\begin{gathered}
\underset{\overline{x}}{\text{minimize}}\;
\sum_{t=i}^{T} \frac{C_t}{k}\Big[\ln(\frac{a}{a-\overline{x}_t-\overline{s}_t})-s_t\Big] \;\;\;\;\;\;\;\;\;\;\\
\text{subject to:}\;\;
\sum_{t=i}^{T}{(\overline{x}_t+{\hat{\overline{s}}_t)}} \geq \frac{\overline{D}}{m}-\sum_{t=1}^{i-1}{(\overline{x}_t+\overline{s}_t)},
\\\overline{x}_t\geq 0;\:\;  ~~~~ t=i,i+1,\cdots,T,\;\;\;\;\;\;\;\;\;\;\;\;\;\;
\\\overline{x}_t\leq \overline{U}_{LED} ;\:\; \;\;\;\; t=i,i+1,\cdots,T,\;\;\;\;\;\;
\end{gathered}
\end{equation}
where $\hat{\bar{s}}_{t}$ represents the predictions of BNN at time $t$. The optimization problem \eqref{optprobedited} is solved once before sunrise and once after sunset. Throughout the day, \eqref{optprobedited} is solved repeatedly at each time step considering sunlight prediction. For simulation, CVXPY is used which enables solving convex optimization problems with high-level features of Python \cite{cvxpy,cvxpy_rewriting}\footnote{CVXPY is available at \url{http://www.cvxpy.org}.}. Interested readers are referred to our previous work \cite{mosharafian2021optimal} for more details on how to calculate the optimal lighting strategy.

%%%%% %%% %%%%%%%%%%%%%%%%%%%%%%%%%%%%%%%%%%%
%%%%% %%% %%%%%%%%%%%%%%%%%%%%%%%%%%%%%%%%%%%
%%%%% %%% %%%%%%%%%%%%%%%%%%%%%%%%%%%%%%%%%%%

\section{Results and Discussion}

\noindent We evaluate the performance of our proposed approach through simulation studies. For each month, the BNN prediction model is trained with twelve years data points and then tested on four years data points. To measure the error rate of the predictions, we use two statistical metrics: Coefficient of Determination ($R^{2}$) and Root Mean Square Error (RMSE) which are calculated as follows:

\begin{equation} \label{R2}
\begin{gathered}
R^{2}=1-\frac{\sum_{i=1}^{N}\,(y_{i}-\hat{y}_{i})^{2}}{\sum_{i=1}^{N}\, (y_{i}-\overline{y})^{2}},\\
RMSE = \sqrt{\frac{\sum_{i=1}^{N}\,(y_{i}-\hat{y}_{i})^{2}}{N}},
\end{gathered}
\end{equation}

\noindent where $y_{i}$ is the observed value, $\hat{y}_{i}$ is the predicted value, $\overline{y}$ is the mean of the observed values, and $N$ is the number of observations. 

\begin{figure}[h!]
  \centering
    \includegraphics[width=1.1\linewidth]{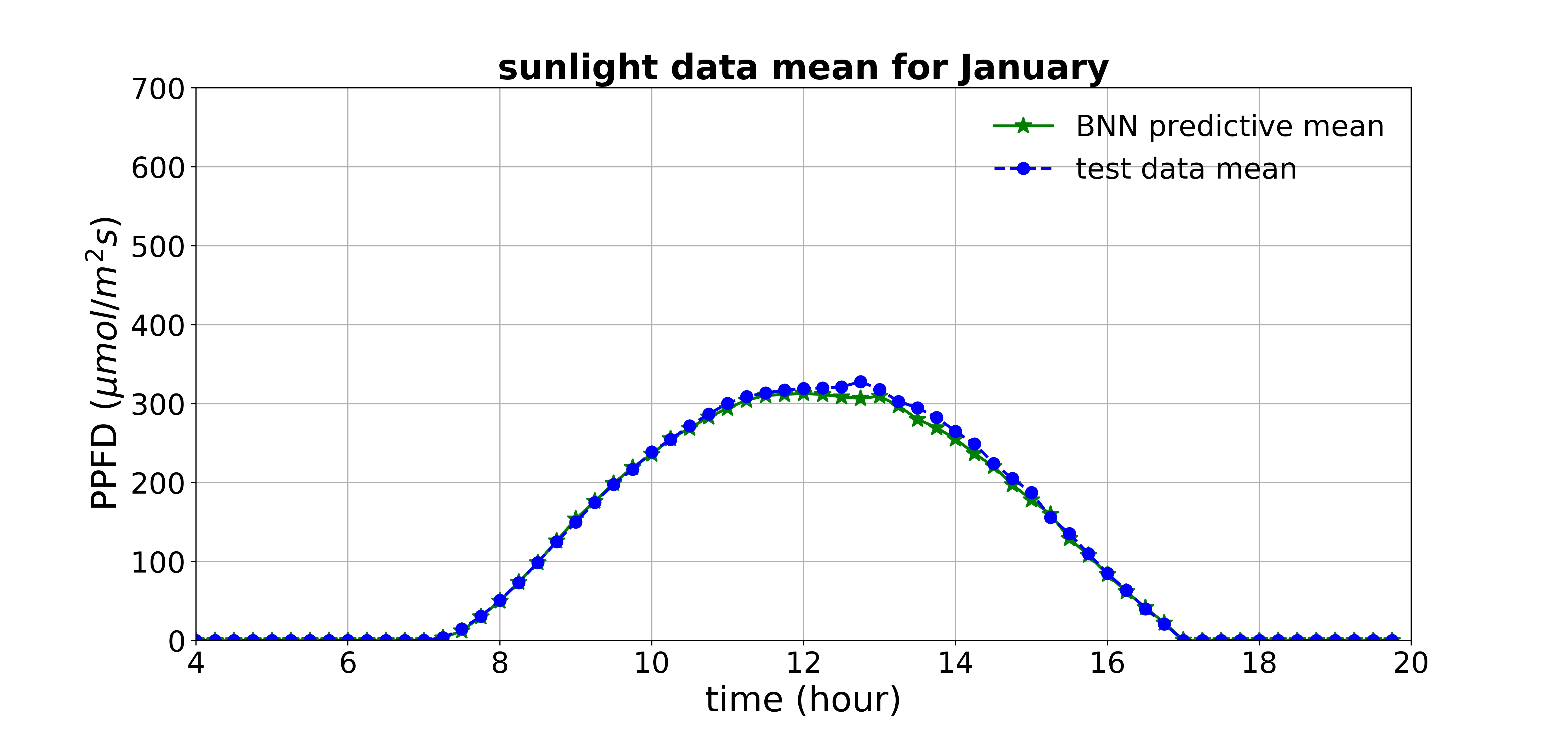}
  \caption{The mean of the actual sunlight values and BNN predicted values for January. $R^{2}=0.998,\,RMSE=5.3\,(1.61\%)$}
  \label{mean_January}
  \vspace{-2mm}
\end{figure}

\begin{figure}[h!]
  \centering
    \includegraphics[width=1.1\linewidth]{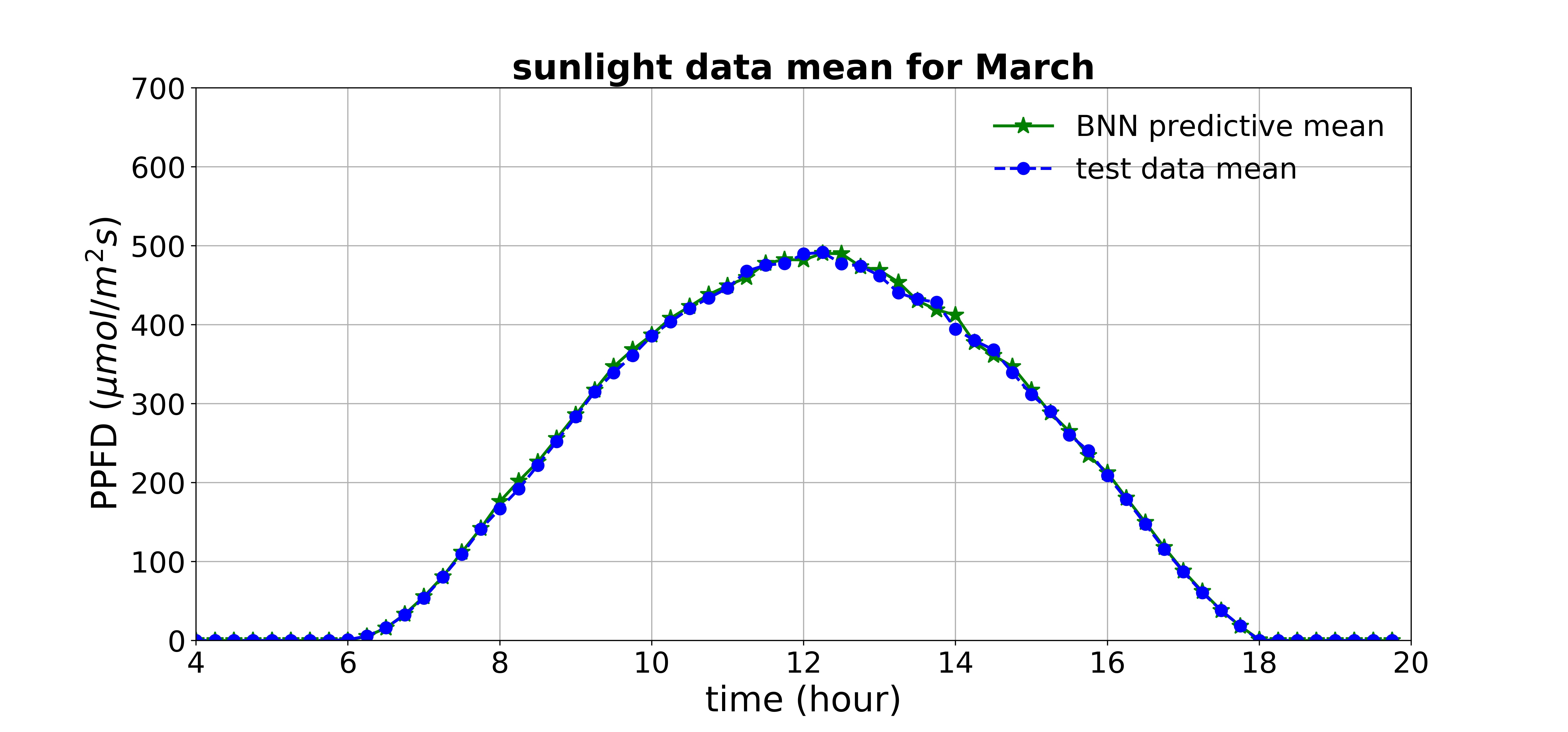}
  \caption{The mean of the actual sunlight values and BNN predicted values for March. $R^{2}=0.999,\,RMSE=5\,(1\%)$}
  \label{mean_March}
\end{figure}

\begin{figure}[h!]
  \centering
    \includegraphics[width=1.1\linewidth]{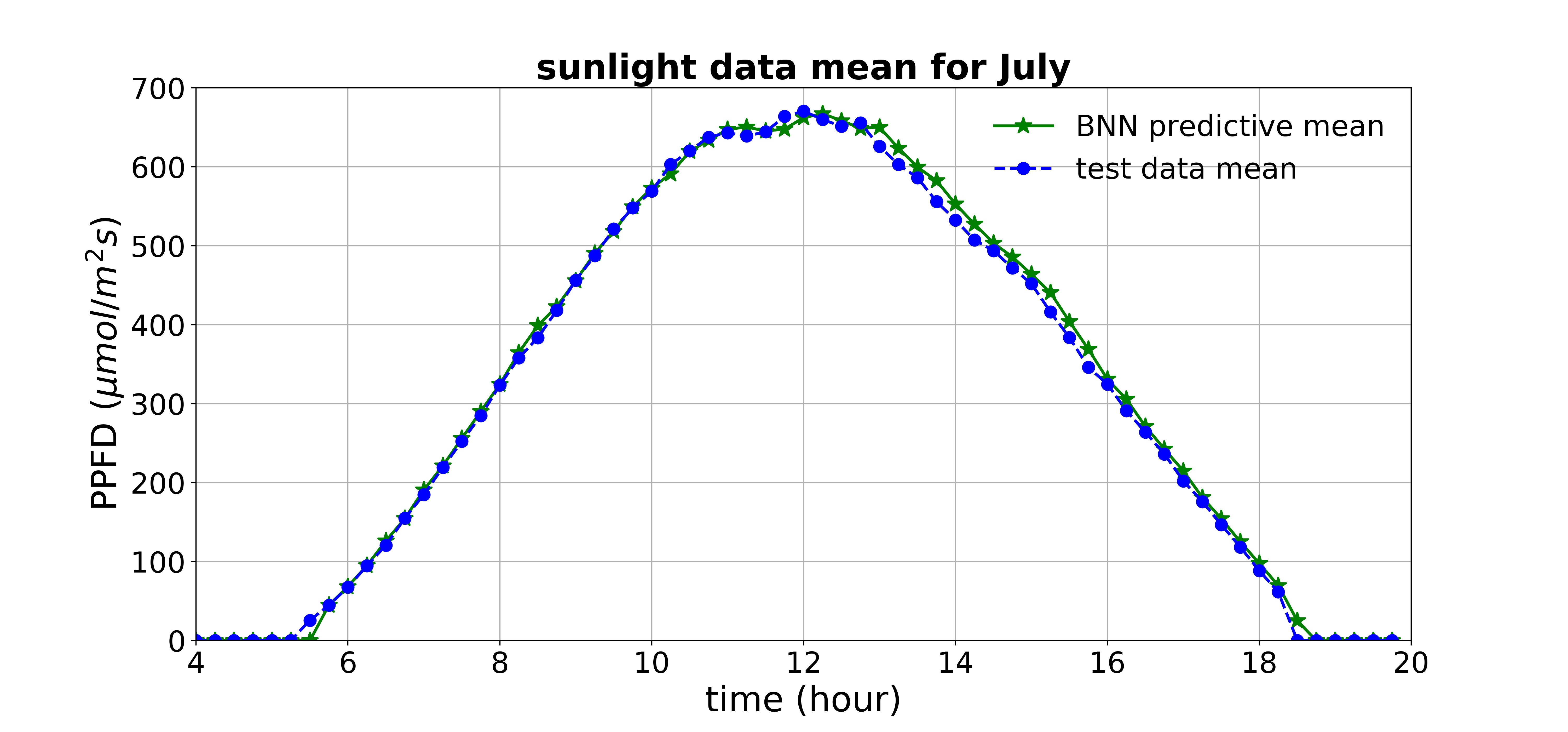}
  \caption{The mean of the actual sunlight values and BNN predicted values for July. $R^{2}=0.997,\,RMSE=11.2\,(1.67\%)$}
  \label{mean_July}
\end{figure}

The BNN model was applied to the test data (four years data), and %then the predicted values were averaged and compared to the average of the test data points. 
then the metrics were computed. $R^{2}$ for all months on average was 0.9971, and RMSE was 9 $\mu\,mol\,m^{-2}\,s^{-1}$ (1.8\%), thereby validating the accuracy of our predictive model. The BNN predictive mean and test data mean for three different months are shown in Fig. \ref{mean_January} to Fig. \ref{mean_July}. Fig. \ref{mean_January} represents a month with low sunlight levels (January), Fig. \ref{mean_March} indicates a month with moderate sunlight levels (March), and Fig. \ref{mean_July} shows a month with high sunlight levels (July).

\begin{figure}[h!]
  \centering
  \includegraphics[width=1.1\linewidth]{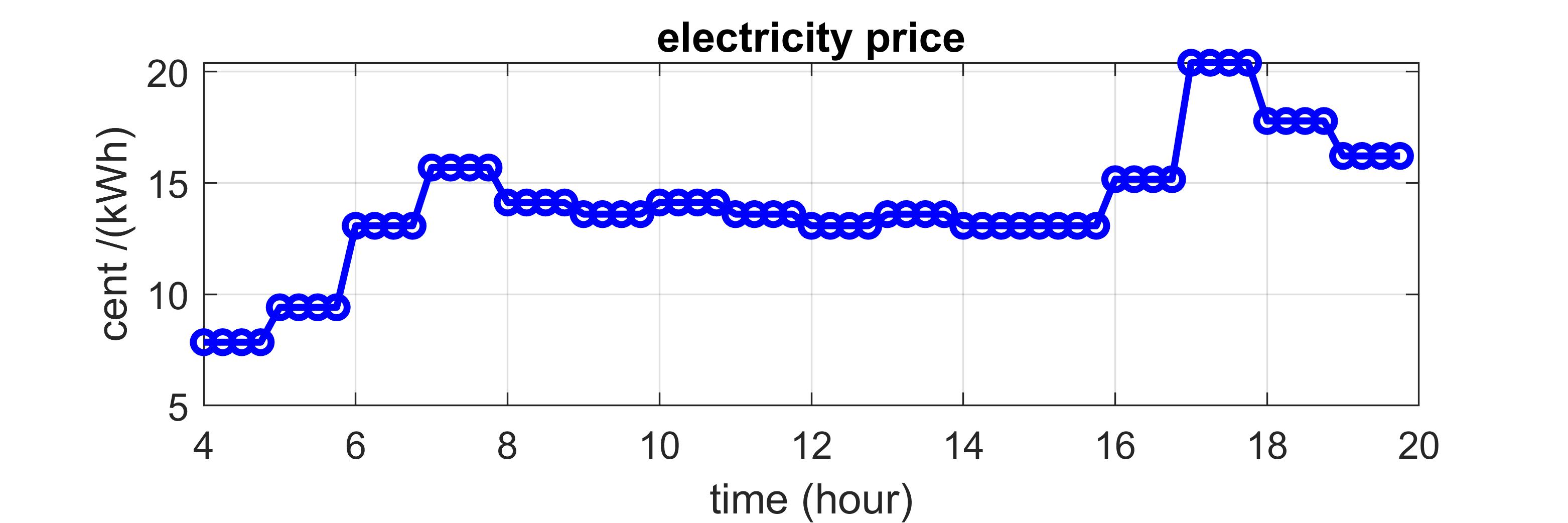}
  \caption{Variable electricity price in cent/\,$kWh$.}
  \label{price}
  \vspace{-4mm}
\end{figure}

After evaluating the performance of our BNN model for sunlight prediction, we solved the optimization problem \eqref{optprobedited} with BNN prediction and compared it to three lighting strategies. The \underline {Markov prediction-based} approach, which uses the Markov model for sunlight prediction to solve the optimization problem \eqref{optprobedited}; the \underline{baseline} method, which is ideal and solves the optimization problem \eqref{optprobedited} assuming perfect prior knowledge of sunlight throughout the day (which is not realistic and only shows a theoretical optimal scenario); and the \underline{heuristic} method, which provides enough supplemental light to reach a constant PPFD, unless the PPFD from sunlight alone exceeds the required level. Thus, by the end of photoperiod, the plants will have received enough supplemental light to reach a DPI greater than or equal to the recommended value. For both prediction-based methods and the baseline, the variable electricity price profile shown in Fig. \ref{price} is used. Heuristic method does not take electricity price or sunlight prediction into account. Optimization parameters used in this work are considered based on \cite{mosharafian2021optimal} and given in Table \ref{param}.

\begin{table}[h!]
\caption{Model and optimization parameters.}
\centering
\renewcommand{\arraystretch}{1.5}
\begin{tabular}{||p{0.85cm}|p{2.6cm}||p{0.85cm}|p{1.5cm}||} 
 \hline\hline
 variable & value & variable & value\\ [0.5ex] 
 \hline\hline
 $a$ & $121 ~\mu\,mol\, m^{-2}\,s^{-1}$
 & $m$ & 900 sec. \\ \hline
 $\overline{D}$ & $3 ~\,mol\, m^{-2}\,d^{-1}$ & $T$ & 64\\\hline
 $\overline{U}_{LED}$ &  $51.47 ~\mu\,mol\, m^{-2}\,s^{-1}$ & $k$ & 0.00277 \\ [1ex] 
 \hline\hline
\end{tabular}
\label{param}
\vspace{-2mm}
\end{table}

\begin{remark} 
To quantify the optimization cost in terms of $cent/kWh$, there is a need to find a conversion factor. In \eqref{optprob}, the unit of $f$ is $cent\, (kWh)^{-1}\, \mu mol \, m^{-2}\, s^{-1}$. First, a factor is needed to convert LED PPFD to $kW \, m^{-2}$. Based on \cite{kusuma2020physics}, $2.8 \,\mu mol \, J^{-1}$ (or equivalently $2.8\times 10^{3}\,\mu mol \, (kW)^{-1} \, s^{-1}$) for LEDs can be used. Finally, time step length (here, $15~min$ or $0.25~ h$) should be considered to reach the unit of $cent\,(kW)^{-1}$. Hence, the conversion factor $l$ is determined to be 
\begin{equation} \label{conversionco}
    \begin{gathered}
        l=\frac{1}{2.8\times 10^{3}}\times 0.25.
    \end{gathered}
\end{equation} 
Consequently, supplemental lighting cost in terms of $cent/m^{2}$ is calculated as $l\times f$.
\end{remark}

\vspace{12pt}

\begin{figure}[h!]
  \centering
    \includegraphics[trim={0 3cm 0 3cm},clip,width=1\linewidth]{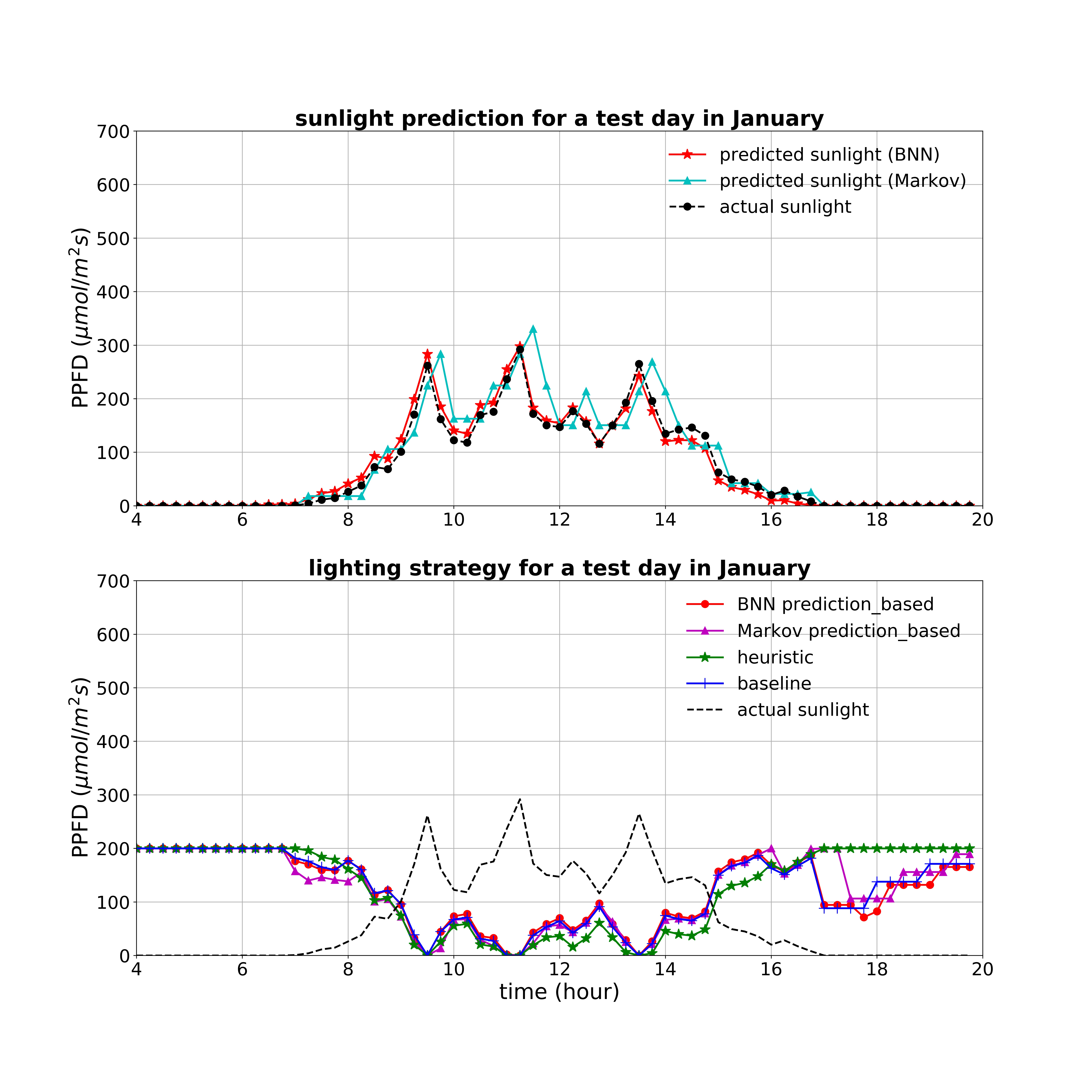}
  \caption{Performance of different predictive strategies for a test day in January. In the first subplot, two approaches for sunlight prediction are compared to the actual sunlight data, and the second subplot shows four lighting strategies: baseline approach, BNN prediction-based method, Markov prediction-based method, and a heuristic approach.}
  \label{Jan}
  \vspace{-4mm}
\end{figure}

\begin{figure}[h!]
  \centering
    \includegraphics[trim={0 3cm 0 3cm},clip,width=1\linewidth]{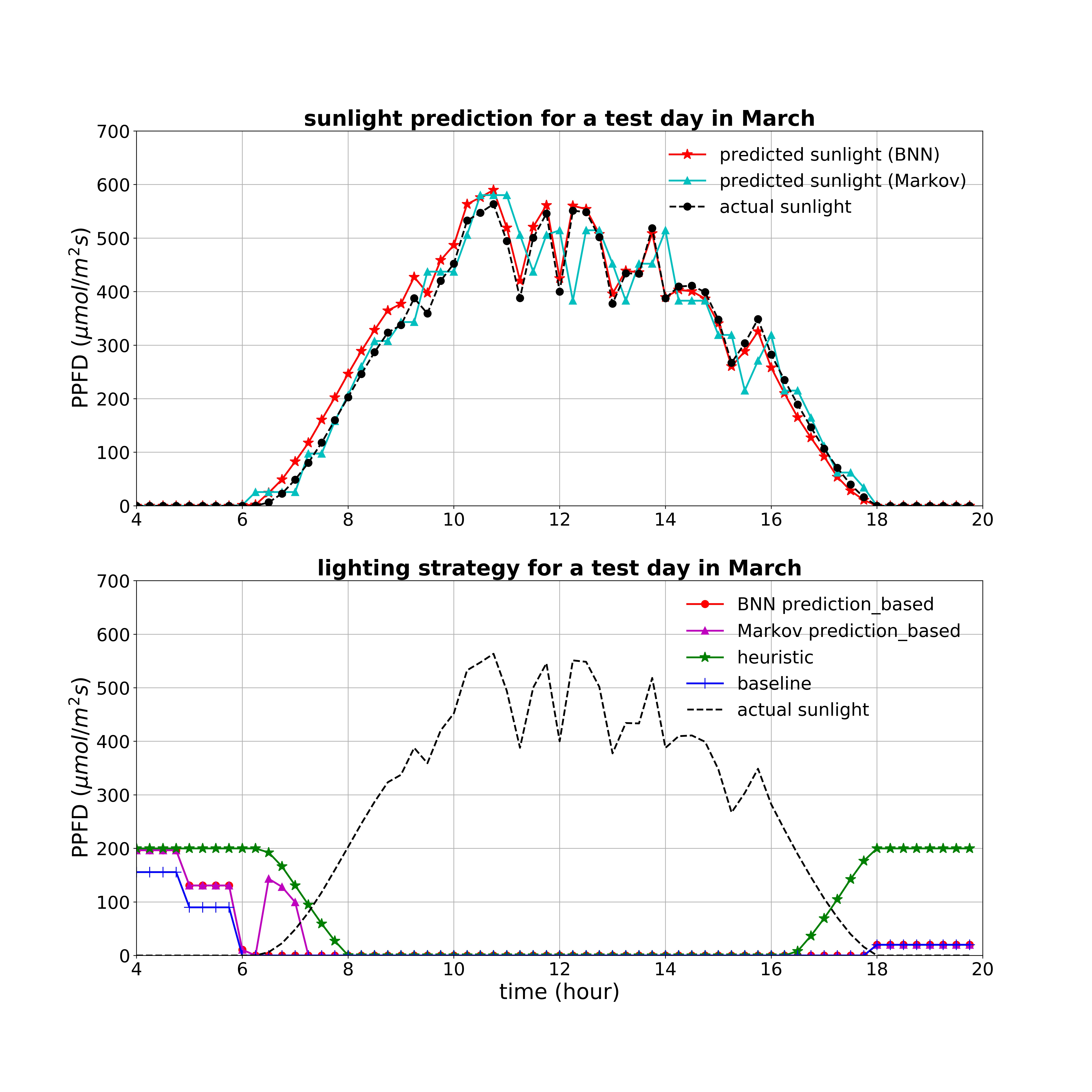}
  \caption{Performance of different predictive strategies for a test day in March.}
  \label{March}
\end{figure}

\begin{figure}[h!]
  \centering
    \includegraphics[trim={0 3cm 0 3cm},clip,width=1\linewidth]{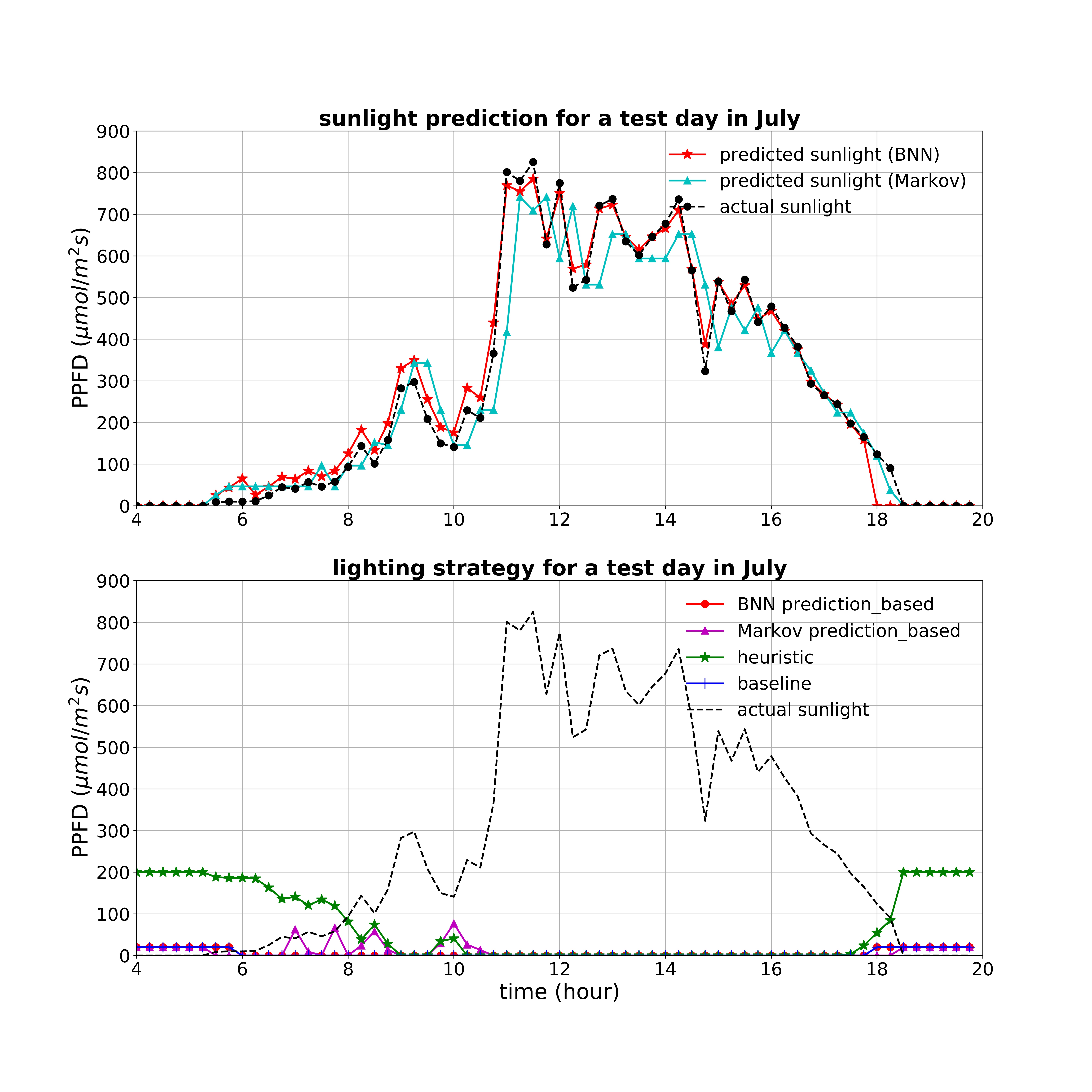}
  \caption{Performance of different predictive strategies for a test day in July.}
  \label{July}
\end{figure}

Three different test days in each month are studied to evaluate the performance of the BNN prediction-based method throughout the year. In Fig. \ref{Jan} -- Fig. \ref{July}, lighting strategies for three test days in three months (January, March, and July) are shown. Besides, actual sunlight and its predictions are also compared. The $R^{2}$ and RMSE of the two prediction methods (BNN and Markov) are compared in Table \ref{prediction}. The lighting cost of the BNN prediction-based method is compared to each of the other methods and shown in Table \ref{cost}.

\vspace{-2mm}

\begin{table} [h!]
\caption{Performance of sunlight predictive models for different test days.}
\centering
\renewcommand{\arraystretch}{1.4}
\begin{tabular}{||p{2.3cm}||p{0.8cm}||p{0.8cm}||p{1cm}||p{1cm}||} 
\hline\hline
Test day & BNN $R^{2}$ & Markov $R^{2}$ & BNN RMSE & Markov RMSE\\ [0.5ex]
\hline\hline
Test day in January & 0.98 & 0.81 & 12.7 (4.23\%) & 35.9 (11.97\%)\\ [1ex] 
\hline
Test day in March & 0.99 & 0.95 & 22.4 (3.93\%) & 46 (8.1\%)\\ [1ex] 
\hline
Test day in July & 0.98 & 0.89 & 32.6 (3.98\%) & 87 (10.61\%)\\ [1ex] 
\hline\hline
\end{tabular}
\label{prediction}
\end{table}

\vspace{3mm}

\begin{table}[h!]
\caption{Cost increase of different lighting strategies (compared to the baseline) in cent$/m^{2}$ per day for different test days.}
\centering
\renewcommand{\arraystretch}{1.5}
\begin{tabular}{||p{2.3cm}||p{1.2cm}||p{1.2cm}||p{1.2cm}||} 
\hline\hline
Test day & BNN prediction-based method & Markov prediction-based method & Heuristic method\\ [0.5ex]
\hline\hline
Test day in January & 0.02 (0.14\%) & 0.09 (0.87\%) & 0.4 (3.86\%)\\ [1ex] 
\hline
Test day in March & 0.26 (26.93\%) & 0.76 (77.42\%) & 5.31 (540.69\%)\\ [1ex] 
\hline
Test day in July & 0 (0\%) & 0.97 (148.58\%) & 5.25 (1346.74\%)\\ [1ex] 
\hline\hline
\end{tabular}
\label{cost}
\end{table}

Based on our previous work \cite{mosharafian2021optimal}, the Markov prediction-based method reduces cost significantly compared to the heuristic method. The results in Table \ref{cost} validate the same conclusion; furthermore, they show that the BNN prediction-based method outperforms even the Markov-based one and reduces cost by up to 59\%, compared to the Markov prediction-based method.

\begin{figure}[h!]
  \centering
    \includegraphics[width=1.1\linewidth]{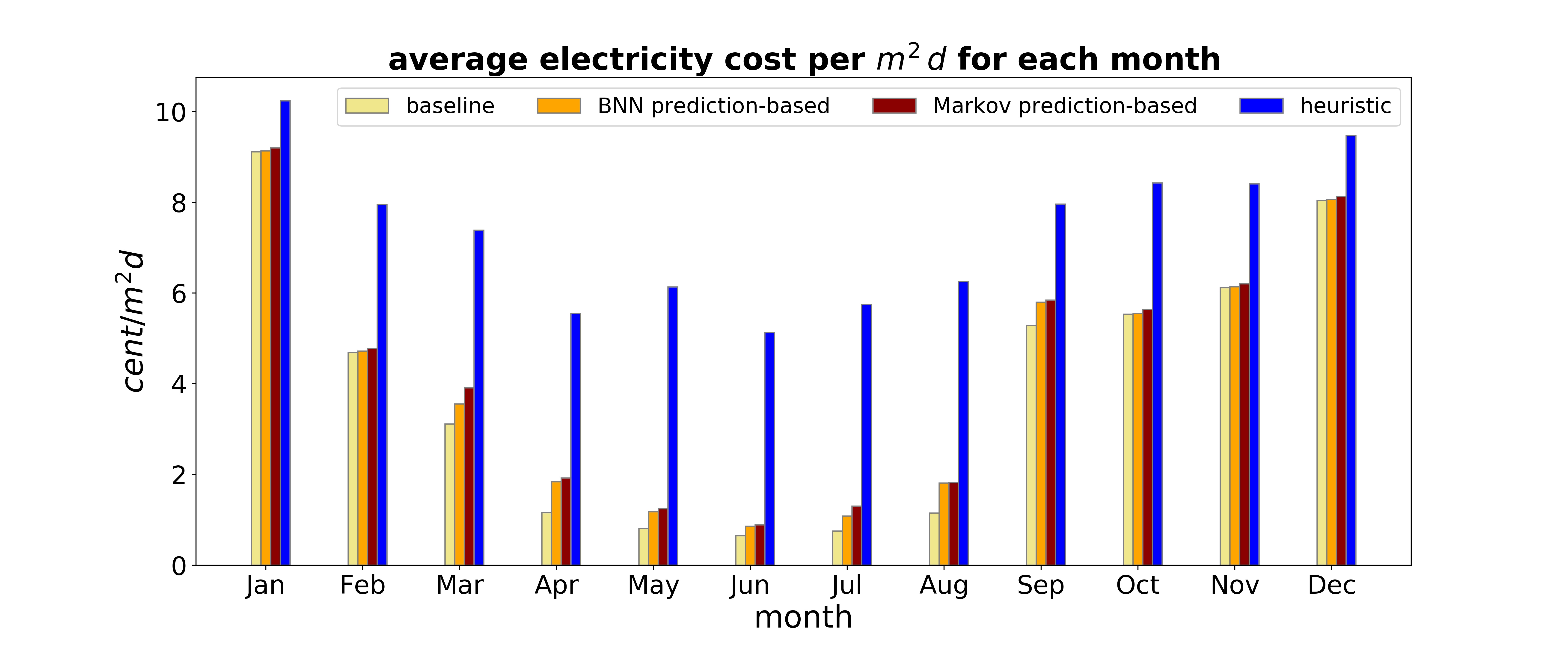}
  \caption{Electricity cost of lighting in $cent/m^{2}/day$ for each month of the year.}
  \label{all_year}
  \vspace{-6mm}
\end{figure}

The performance of the BNN prediction-based method is compared to the other three methods for a whole year in Fig. \ref{all_year}. For each month, the cost of different lighting strategies was averaged over three days. Sunlight levels were much lower during the winter and fall months compared to the spring and summer; therefore, as shown in Fig. \ref{all_year} more supplemental light was provided to reach the minimum DPI during the cold months, which results in higher electricity cost (for all approaches). For instance, for the test day in January (Fig. \ref{Jan}), the sunlight levels were low, and that results in providing much supplemental light to satisfy the light requirements of the plants with all lighting approaches. Thus, there is not a huge difference in the cost of prediction-based methods and the heuristic method. The results in Fig. \ref{all_year} show that, on average, the BNN prediction-based method brings about 2.27\% and 43.91\% cost reduction compared to the Markov prediction-based method and the heuristic method, respectively.   

%%%%%%%%%%%%%%%%%%%%%%%%%%%%%%%%%%%%%%%%%%%%%%%%%%%%%%%%%%%%%%%%%%

\section{Conclusion}

In this paper, a new method was proposed to find an optimal supplemental lighting strategy in greenhouses equipped with LEDs. The problem was formulated as a convex optimization problem aiming at minimizing supplemental lighting cost while providing a minimum light level needed by plants. The proposed algorithm considered a BNN model for predicting sunlight intensities, whose performance was evaluated through statistical metrics and showed high accuracy. The electricity cost of the proposed method was compared to a Markov-based method and a heuristic control strategy and brought about improvement in cost savings. 

%%%%%%%%%%%%%%%%%%%%%%%%%%%%%%%%%%%%%%%%%%%%%%%%%%%%%%%%%%%%%%%%%%%%%

\bibliographystyle{ieeetr}
\bibliography{references}

\end{document}